\documentstyle[preprint,aps]{revtex}
\def\d{{\rm d}}
\begin{document}
\draft
\title{Fluctuation-Induced Interactions between Rods on Membranes and
Interfaces}
\author{Ramin Golestanian}
\address{Institute for Advanced Studies in Basic Sciences,
Zanjan   45195-159, Iran}
\author{Mark Goulian\footnote{Address after Aug. 31:
{\it Center for Studies in Physics and Biology,
The Rockefeller University, 1230 York Avenue,
New York, NY 10021.}}}
\address{Exxon Research and Engineering,
Annandale, NJ  08801}
\author{Mehran Kardar}
\address{Department of Physics, Massachusetts Institute of
Technology, Cambridge, MA 02139}
\date{\today}
\maketitle
\begin{abstract}
We consider the interaction between two rods embedded in a
fluctuating surface which is governed by either surface tension
or rigidity. The modification of fluctuations by the rods
leads to an attractive long-range interaction
that falls off as $1/R^4$ with their separation.
The orientational dependence of the resulting
interaction is non-trivial and may lead to interesting patterns
of rod-like objects on such surfaces.
\end{abstract}
%%
%\begin{multicols}{2}
%%

In both cell biology and in the design of biomolecular materials
it is important to understand the interactions between inclusions
in fluid membranes. The lipid
bilayer defines the outer boundaries of both the cell and its interior
organelles and vesicles. However, biomembranes play far more than simply a
structural role; they also regulate and act as host for many  bio- chemical
and
physical processes \cite{alb,gen}. For example, they control and regulate
inter-
and intra-cellular recognition and transport,
adhesion, ion concentrations, and energy conversion. It is not surprising
then, that biomembranes are far from homogeneous.
In addition to the many different lipids that make up the bilayer,
biomembranes contain a variety of proteins, glycolipids,
and other macromolecules.
Inclusions have also been incorporated in artificial
membranes \cite{gen,shen}.
Such model-membrane systems have potential applications
for targeted drug delivery and may also lead to novel ``biologically
inspired'' materials, such as nano-scale pumps, templates,
functionalized interfaces, and chemical reactors. Thus, for both
biological and artificial membranes, it is important to understand
how inclusions affect the physical properties of the membrane,
and how the membrane in turn contributes to the interactions between
inclusions.

There are a number of different forces which act within
membranes \cite{isr}:
van der Waals interactions fall off with separation $R$, as $1/R^6$ at long
distances; the Coulomb interaction is strongly screened under physiological
conditions;
%(typical ion concentrations are a few milimolar, giving a
%screening length of less than 10\AA)
hydration and structural forces are also short-ranged. There are additional
interactions
between inclusions which are mediated by the membrane;
the inclusion disturbs the lipid bilayer and this disturbance propagates
to neighboring inclusions (c.f. \cite{gen,isr,mou,goul,dan} and references
therein). When macroscopic thermal fluctuations are unimportant (we
refer to this case as $T=0$), the resulting interactions tend to be
short-ranged, falling off exponentially with a characteristic length
corresponding to the distance over which the lipid disturbance ``heals.''
For example, if in the region around an inclusion the membrane is
forced to deviate from its preferred thickness ($\sim 40$\AA), then
the resulting disturbance decays exponentially with a length
comparable to this thickness \cite{dan}.

If the membrane is to mediate long-ranged interactions, then in the
long-distance limit it should be possible to neglect the membrane
thickness and details concerning lipid structure. In this limit,
the membrane is well-described by the elastic Hamiltonian \cite{CH},
\begin{equation}
{\cal H}=\int \d S \left[\sigma+\frac{\kappa}{2} H^2+\bar\kappa
K \right],\label{CHH}
\end{equation}
where $\d S$ is the surface area element, and $H$, $K$ are the mean and
Gaussian curvatures respectively. The elastic properties of the surface
are described by the tension $\sigma$, and the bending rigidities
$\kappa$ and $\bar\kappa$. A finite surface tension is the
most
important coupling in Eq.(\ref{CHH}) and dominates the bending terms at
long wavelengths. This is the case for films on a frame, interfaces at
short distances, and possibly closed membranes in the presence of osmotic
pressure  differences between their interior and exterior. On the other
hand,
for closed bilayers in the absence of osmotic stress, as well as for
microemulsions, the surface tension is effectively zero \cite{DGT,BL,DL}.
In these cases, the energy cost of fluctuations is controlled by the
rigidity
terms. For simplicity we shall refer to surface tension dominated surfaces
as
films, and to rigidity controlled ones as membranes.

The long distance interactions between inclusions in a membrane were
examined in
Ref. \cite{goul}.  It was shown that if the inclusions are asymmetric across
the bilayer and impose a local curvature, Eq.(\ref{CHH}) gives rise
to a repulsive ($T=0$) interaction that is long-ranged, falling off with
distance as $1/R^4$. The energy scale of the interaction is set by
$\kappa$ and $\bar\kappa$. If thermal fluctuations of the membrane are
included ($T\ne0$), on the other hand, there is a $1/R^4$ interaction for
{\it generic} inclusions, as long as the rigidity of the inclusion differs
from that of the ambient membrane \cite{goul}. In particular,
if the inclusions are much stiffer than the membrane, the potential has the
form
\begin{equation}
V(R)=-k_BT\frac{6A^2}{\pi^2 R^4},
\end{equation}
where $A$ is the area of each inclusion. (In the above formula,
the result in Ref. \cite{goul} has been corrected by a factor of
$1/2$ \cite{GGK}.)
The interaction is  attractive and independent of $\kappa$ and
$\bar\kappa$; the
energy scale  is set by $k_BT$.

In this letter we consider the orientational interaction between
{\it rod-like} inclusions due to thermal
surface fluctuations ($T\ne 0$). The rods are assumed to be much more
rigid than the membrane so that they cannot move coherently with
it; the only degrees of freedom for the rods
are rigid translations and rotations while they remain attached to the
surface. Conversely, the rods impose rigid boundary conditions
on the surface, constraining its shape fluctuations.
We will first give the results for the interaction energy and then
briefly sketch their derivation; details of the calculations will be
described in a separate paper \cite{GGK}. Finally, we discuss some
general aspects of this orientational interaction.

Consider a surface containing two rods of lengths
$L_1$ and $L_2$ separated by a distance $R\gg L_i$. The rods make angles
 $\theta_1$ and $\theta_2$, respectively, with the line connecting their
center of masses (see Fig.~1). For fluctuating films, we find
an attractive fluctuation-induced interaction given by,
\begin{equation}\label{film}
V_F(R,\theta_1,\theta_2)=-\frac{k_{B}T}{128}
\frac{L_1^2 L_2^2}{R^4}\cos^2\left[\theta_{1}+\theta_{2}\right]
+O\left(1/R^6\right).
\end{equation}
Note that this behavior is the {\it square} of a dipole-dipole
interaction in two dimensions, with $L_1$ and $L_2$ as the dipole
strengths.
The fluctuation-induced interaction on a membrane is very
similar
and given by
\begin{equation}\label{memb}
V_M(R,\theta_1,\theta_2)=-\frac{k_{B}T}{128}
\frac{L_1^2 L_2^2}{R^4}\cos^2\left[2\left(\theta_{1}+\theta_{2}
\right)\right]+O\left(1/R^6\right).
\end{equation}
In this case the interaction has the surprising property of being minimized
for both parallel and perpendicular orientations of the rods.
These interactions should have important effects in aligning asymmetric
inclusions in biomembranes. Since orientational
correlations are often easier to measure than forces, this result may
also be useful in probing the fluctuation-induced interactions.
Finally, this interaction may lead to novel two-dimensional
structures for collections of rod-like molecules. In particular, the
resemblance of the orientational part of the interaction to dipolar forces
suggests that a suitable way to minimize the energy of a collection of rods
is to
form them into chains. (If the rods are not colinear, the interactions
cannot be
minimized simultaneously.) Such chain like structures are observed for
ferromagnetic particles controlled by similar forces\cite{ros}.

To obtain Eq.(\ref{memb}) we start with a thermally fluctuating planar
membrane subject to the Hamiltonian in Eq.(\ref{CHH}) (with $\sigma=0$). We
assume that the size of the membrane $d$, is well below the persistence
length
$\xi$\cite{DGT}. In this limit, the membrane experiences only small
fluctuations
about a flat state. We may then parametrize the membrane surface with a
height
function $u(r)$, and approximate the full coordinate-invariant Hamiltonian
in Eq.(\ref{CHH}) by the Gaussian form
${\cal H}_0=\kappa/2 \int\d^2r
\left(\nabla^{2}u(r)\right)^2.$
Now consider the situation depicted in Fig.~1, where two rigid,
rod-like objects, are attached to the membrane.
We shall represent the rods by narrow rectangles
of lengths $L_1$ and $L_2$, and widths $\epsilon_1$ and
$\epsilon_2$; ultimately taking the limit of $\epsilon_i\to0$.
The rods are constrained to
fluctuate with the membrane but, due to their stiffness, can only
be tilted or translated up and down rigidly. We can parametrize
all possible configurations of the rods by
\begin{eqnarray}
{\hskip 4cm} u(r)|_{r\in L_i}=a_i+{\bf{b}}_i\cdot{\bf{r}},
{\hskip 1.5cm}{\rm for}\quad i=1,2\qquad,  \label{abdef}
\end{eqnarray}
where we have also used $L_i$ to denote the $i$th rod.

To calculate the partition function, we follow a procedure similar
to Ref.\cite{goul} and sum over all possible
configurations of the membrane surface,
weighted by the corresponding Boltzmann weight, and subject to the
constraints imposed by the rods via Eq.(\ref{abdef}). The constraints
may be implemented with the aid of delta functions as
in Ref. \cite{LiK}, leading to
\begin{equation}
{\cal Z}=\int {\cal D}u(r) \prod_{i=1}^{2} \int \d a_i
\d^2b_i
	\prod_{r'\in{L_i}}
\delta\left(u(r')-a_i-{\bf{b}}_i\cdot {\bf{r}}'\right)
	\exp\left[-\frac{{\cal H}_0}{k_{B}T}\right]. \label{Z1}
\end{equation}
In Eq.(\ref{Z1}) we have included only the leading term in an expansion
in powers of ${\bf b}_i$, neglecting higher order terms
from the projection of $L_i$ onto the $x-y$ plane, as well as from the
integration measure for ${\bf b}_i$ on the sphere of unit normals.
Since ${\bf b}_i$ controls the gradient of $u(r)$ at the boundary of $L_i$,
the expansion in ${\bf b}_i$ is in the same spirit as the gradient
expansion for ${\cal H}_0$. It can be shown that these higher
order terms are irrelevant in the  limit $d\ll\xi$ \cite{GGK}.
Expressing the delta functions as functional integrals over auxiliary
fields $k_i(r)$ defined on the rods, we obtain
\begin{eqnarray}\label{Z2}
{\cal Z}=\int {\cal D}u(r) \prod_{i=1}^2 \int \d a_i
\d^2b_i\int_{L_i}{\cal D}k_i(r)\exp\Biggl[&&-\frac{\kappa}{2k_{B}T}
\int \d^2r \left(\nabla^{2}u(r)\right)^2\\
&&+i \sum_{i=1}^2\int_{L_i}\d^2r_i k_i(r_i)
\left(u(r_i)-a_i-{\bf{b}}_i\cdot{\bf{r}}_i\right)\Biggr].\nonumber
\end{eqnarray}
Integrating out $u(r)$,  $a_i$, and ${\bf b}_i$, then gives
(throughout we drop irrelevant multiplicative constants in ${\cal Z}$)
\begin{eqnarray}\label{coul}
{\cal Z}=\prod_{i}\int{\cal D}k_i(r)\delta\biggl(\int_{L_i}
\d^2r_i&&k_i(r_i)\biggr)\delta\left(\int_{L_i}\d^2r_i{\bf r}k_i(r_i)
\right)\times\\
&&\exp\left[-\frac{k_BT}{2\kappa}\sum_{i,j=1}^2\int_{L_i}\d^2r_i
\int_{L_j}\d^2r_j
k_i( r_i)G({\bf r}_i-{\bf r}_j)k_j(r_j)\right],\nonumber
\end{eqnarray}
where $G({\bf r}-{\bf r}')=(1/8\pi)\,\mid{\bf r}-{\bf r}'\mid^2\ln\mid{\bf
r}-{\bf r}'\mid.$

Equation (\ref{coul}) is analogous to the partition function for a pair of
plasmas confined to the interior of rods $L_1$ and $L_2$. The delta
functions impose the constraints that the net charge and dipole
moments vanish within each rod. When the distance $R$ between rods
is much larger than their size (i.e. $L_i\ll R$), we may approximate
$G({\bf r_1}-{\bf r_2})$ in  Eq.(\ref{coul}) by a multipole expansion
and keep only the leading term, which comes from
the quadrupole moments
$Q^{(i)}_{ab}\equiv\int_{L_i}\d^2r\;r_a r_b\, k_i(r)$, to get
\begin{eqnarray}\label{ZK}
{\cal Z}&=&\prod_{i}\int{\cal D}k_i(r)\int \d{\bf Q}^{(i)}\d a_i
\d^2b_i
d{\bf g}^{(i)}
\;\exp\left\{-\frac{k_BT}{2\kappa}\sum_{i}\int_{L_i}\d^2r\d^2r'k_i(r
)
G({\bf r}-{\bf r'})k_i(r')\right.\nonumber\\
&&-i\sum_{i}\int_{L_i}\d^2r k_i(r)\left[a_i+{\bf b}_i\cdot{\bf r}+
{\bf r}\cdot{\bf g}^{(i)}\cdot{\bf r}\right]
\left. +i\sum_{i}g^{(i)}_{ab}Q^{(i)}_{ab}-\frac{k_BT}{2\kappa}
v\left[{\bf Q}^{(1)},{\bf Q}^{(2)}\right]\right\},
\end{eqnarray}
with the quadrupole-quadrupole interaction $v\left[{\bf Q}^{(1)},
{\bf Q}^{(2)}\right]$ the same as in Ref.\cite{goul}.

We first isolate the integration over $k_1(r)$ in Eq.(\ref{ZK}),
\begin{eqnarray}
I_1\equiv\int{\cal D}k_1(r)\d a_1\d^2b_1\exp\Biggl\{&&-\frac{k_BT}{2\kappa}
\int_{L_1}\d^2r\d^2r'
k_1(r)G({\bf r}-{\bf r}')k_1(r')\\
&&-i\int_{L_1}\d^2r k_1(r) \left[a_1+{\bf
b}_1\cdot{\bf r}+
{\bf r}\cdot{\bf g}^{(1)}\cdot{\bf r}\right]\Biggr\}.\nonumber\label{Idef}
\end{eqnarray}
To perform the above integration, the Green's function $G({\bf r}-{\bf
r}')$ has to be inverted in the finite region $L_1$. To do
this, we introduce an auxiliary field $h({\bf r})$ over the whole plane
(indicated by ${\rm IR}^2$) and write
\begin{equation}
I_1=\int{\cal D}h(r)\d a_1\d^2b_1\exp\left[-\frac{\kappa}{2k_BT}
\int_{{\rm IR}^2}\d^2r
\left(\nabla^2 h(r)\right)^2\right]
\prod_{r'\in L_1}\delta\left(h(r')-a_1-{\bf b}_1\cdot{\bf r}-
{\bf r}\cdot{\bf g}^{(1)}\cdot{\bf r}\right).
\end{equation}
Integration over $h(r),a_1,{\bf b}_1$ gives (dropping a multiplicative
constant)\cite{GGK}
\begin{equation}
I_1=\exp\left\{-\frac{\kappa}{k_BT}\left[2\epsilon_1L_1\,
\left(g^{(1)}_{aa}\right)^2 +\pi\left(L_1
g^{(1)}_{xy}\right)^2\right]\right\}.
\label{I}
\end{equation}

The result of the $k_2(r)$ integration in Eq.(\ref{ZK}) is similar,
with the index 2 replacing 1, and with the coordinate axis
appropriately rotated to align with the second rod.
The overall expression for the partition function now reads
\begin{eqnarray}
{\cal Z}=\prod_{i=1}^2 \int \d{\bf Q}^{(i)} \d{\bf g}^{(i)}
\exp\Biggl\{-\frac{\pi\kappa}{k_BT}&&\left[\left(L_1\;g^{(1)}_{x'y'}\right)^2
+\left(L_2\;g^{(2)}_{x''y''}\right)^2\right]\Biggr\}\times\\
&&\exp\left\{-i\sum_{i} g^{(i)}_{ab} Q^{(i)}_{ab}
-\frac{k_BT}{2\kappa}
v\left[{\bf Q}^{(1)},{\bf Q}^{(2)}\right]\right\} \nonumber,
\end{eqnarray}
where we have set the widths of the rods to zero
($\epsilon_i \to0$). The primed indices $x',y',x'',y''$ indicate
that the corresponding components are with respect to the
coordinate frames where $L_1\parallel y'$ and $L_2\parallel y''$.
We define an un-primed coordinate system such that the $x$-axis
is parallel to $\hat{R}$ and the two rods make angles of $\theta_1$
and $\theta_2$ with respect to the $x$-axis as in Fig.~1.
After performing the remaining integrations, which are cumbersome
but straightforward, we end up with the $(R,\theta_1,\theta_2)$
dependent part of the free energy given in Eq.(\ref{memb}).

The calculation for films is similar. In this case,
the bending rigidity is set to zero in Eq.(\ref{CHH}). We restrict to
sufficiently large surface tensions such that
$\sigma a^2/k_{B}T \gg 1$, with $a$ some microscopic length.
We can then approximate Eq.(\ref{CHH}) by the Gaussian Hamiltonian
${\cal H}_0=\sigma/2\, \int\d^2r\left(\nabla u(r)\right)^2.$
Computing the partition
function along similar lines, we find that
the leading behavior results from fluctuating dipoles instead of
quadrupoles.
However, the calculation is complicated by the fact that the dependence on
${\bf b}_i$ is more complex than in the case of the membrane.
To extract a simple answer, we assume $\sigma L^4/R^2\ll k_BT$,
and find the result in Eq.(\ref{film}).

The interactions in Eqs.(\ref{film}) and (\ref{memb}) have a number of
interesting properties. First, their magnitude is set by $k_BT$ and is
independent of the tension and rigidity coefficients $\sigma$ and $\kappa$.
Thus the effect persists even for rather stiff membranes with
$\kappa\gg k_BT$; as long as the inclusions are more rigid than the
embedding surface.
Second, the interaction falls off with distance as $1/R^4$.
This is a general feature of fluctuation-induced forces,
including the van der Waals interaction, which in $d$ dimensions
falls off as $1/R^{2d}$. (Surprisingly, a recent calculation of similar
orientational forces for a polymer on a flexible surface \cite{Podgornik}
finds different asymptotic decays for films and membranes.) Since the
direct
van der Waals interactions fall off as $1/R^6$, the forces
mediated through the two-dimensional surface will always dominate
asymptotically.
Finally, the most interesting aspect is the orientational
dependence of the force: The angular dependence is a
{\it squared dipolar interaction} for inclusions on a film, and a
{\it squared quadrupolar interaction} on a membrane.
By comparison, an additive interaction, between any
two
infinitesimal line elements, leads to an orientation dependence of
$\cos2\theta_{1}+\cos2\theta_{2}$.
This angular dependence is completely different, and
minimized when the two rods are parallel to their axis of separation.
Presumably both interactions are present for rods of finite thickness;
the additive interaction is proportional to $(L\epsilon)^2$, where
$\epsilon$ is the thickness. The previously calculated interactions
are thus larger by a factor proportional to $(L/\epsilon)^2$ and
should dominate for thin rods.

%\acknowledgements
MK and MG acknowledge the hospitality of the ITP at Santa Barbara
where this work was initiated (supported by NSF Grant No. PHY-89-04035).
The work at MIT is supported by the NSF grant DMR-93-03667. RG
acknowledges support from the Institute for Advanced Studies in Basic
Sciences at Gava Zang, Zanjan, Iran.

\begin{figure}
\caption{The $i$th rod has length $L_i$, width $\epsilon_i$, and makes an
angle
$\theta_i$ with the line joining the centers of the two rods. In the
text, the distance between rods $R$ is taken to be much larger than $L_1$
and
$L_2$.}
\end{figure}


\begin{references}

\bibitem{alb}
 B. Alberts, J. Lewis, M. Raff, K. Roberts and J.D. Watson, {\it Molecular
Biology of the Cell}, Garland, New York 1994.
\bibitem{gen}
 R.B. Gennis, {\it Biomembranes, Molecular Structure and Function},
 Springer-Verlag, New York 1989.
\bibitem{shen}
 Y. Shen, C.R. Safinya, K.J. Rothschild, K.S. Liang, and A.F. Ruppert,
 {\it Nature}, {\bf 366} (1993) 48 (and references therein).
\bibitem{isr}
 J. Israelachvili, {\it Intermolecular and Surface Forces}, Academic Press,
 San Diego 1992.
\bibitem{mou}
 O.G. Mouritsen and M. Bloom, {\it Annu. Rev. Biophys. Biomol. Struct.},
 {\bf 22} (1993) 145.
\bibitem{goul}
 M. Goulian , R. Bruinsma, and P. Pincus, {\it Europhys. Lett.}, {\bf 22}
(1993) 145; Erratum in {\it Europhys. Lett.} {\bf 23} (1993) 155.
 \bibitem{dan}
 N. Dan, P. Pincus, and S.A. Safran, {\it Langmuir}, {\bf 9} (1993) 2768.
\bibitem{CH}
 P.B. Canham, {\it J. Theor. Biol.}, {\bf 26} (1970) 61;
 W. Helfrich, {\it Z. Naturforsch.}, {\bf 28c} (1973) 693.
\bibitem{DGT}
 P.G. de Gennes and C. Taupin, {\it J. Phys. Chem.}, {\bf 86} (1982) 2294.
\bibitem{BL}
 F. Brochard and J.F. Lennon, {\it J. de Phys.}, {\bf 36} (1975) 1035.
\bibitem{DL}
	F. David and S. Leibler, {\it J. de Phys.} II, {\bf 1} (1991) 959.
\bibitem{ros}
 R.E. Rosensweig, {\it Ferrohydrodynamics}, Cambridge, New York 1985.
\bibitem{LiK}
H. Li and M. Kardar, {\it Phys. Rev. Lett.} {\bf 67}, 3275 (1991);
{\it Phys. Rev.} {\bf A46}, 6490 (1992).

\bibitem{Podgornik}
R. Podgornik, preprint, cond-mat/9503146 (1995).

\bibitem{GGK}
R. Golestanian, M. Goulian, M. Kardar, in preparation.
\end{references}
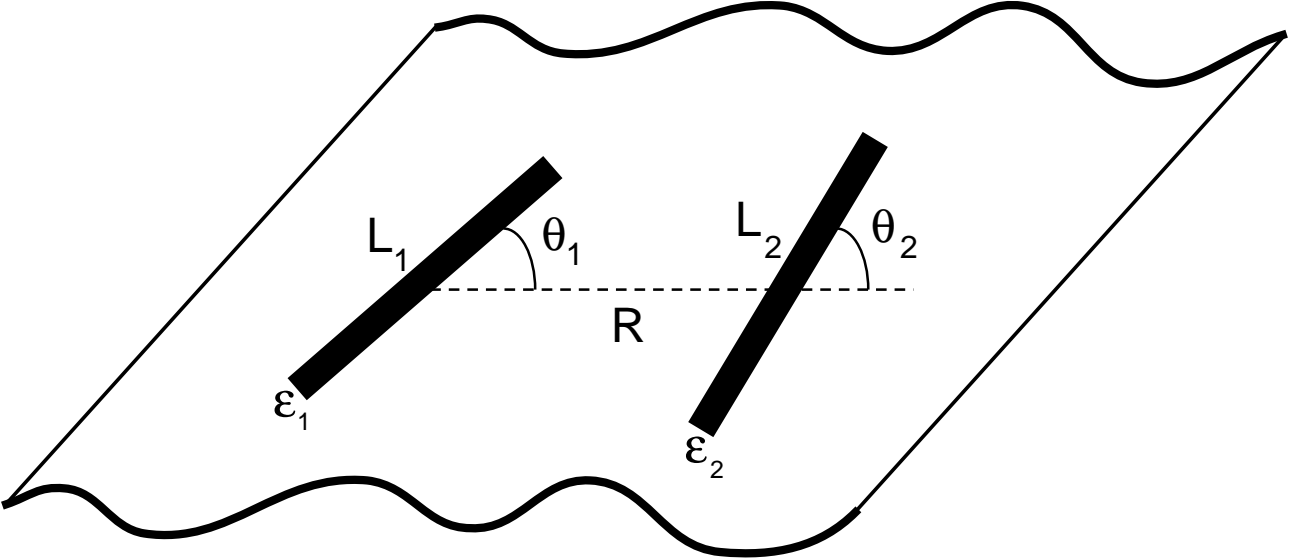
\end{document}